\documentclass[11pt,a4paper,final]{iopart}

\usepackage{color}
\usepackage{amsmath}

\usepackage{iopams}
\usepackage{mathtools}
\usepackage{graphicx}
\usepackage{dcolumn}
\usepackage{array}
\usepackage{lipsum}
\usepackage{bm}
\usepackage{subfigure}
\usepackage{multirow}
\usepackage{tabularx}
\usepackage{braket}
\graphicspath{{plots/}}

\newcommand{\beq}{\begin{equation}}
\newcommand{\eeq}{\end{equation}}
\newcommand{\bea}{\begin{eqnarray}}
\newcommand{\eea}{\end{eqnarray}}

\begin{document}

\title{Strong geometry dependence of the X-ray Thomson Scattering Spectrum in single crystal silicon}

\author{Thomas Gawne}
\address{Center for Advanced Systems Understanding (CASUS), D-02826 G\"orlitz, Germany}
\address{Helmholtz-Zentrum Dresden-Rossendorf, D-01328 Dresden,  Germany}
\ead{t.gawne@hzdr.de}

\author{Zhandos A. Moldabekov}
\address{Center for Advanced Systems Understanding (CASUS), D-02826 G\"orlitz, Germany}
\address{Helmholtz-Zentrum Dresden-Rossendorf, D-01328 Dresden,  Germany}

\author{Oliver S. Humphries}
\address{European XFEL, D-22869 Schenefeld, Germany}

\author{Karen Appel}
\address{European XFEL, D-22869 Schenefeld, Germany}

\author{Carsten Baehtz}
\address{Helmholtz-Zentrum Dresden-Rossendorf, D-01328 Dresden,  Germany}

\author{Victorien Bouffetier}
\address{European XFEL, D-22869 Schenefeld, Germany}

\author{Erik Brambrink}
\address{European XFEL, D-22869 Schenefeld, Germany}

\author{Attila Cangi}
\address{Center for Advanced Systems Understanding (CASUS), D-02826 G\"orlitz, Germany}
\address{Helmholtz-Zentrum Dresden-Rossendorf, D-01328 Dresden,  Germany}

\author{Celine Cr\'episson}
\address{Department of Physics, University of Oxford, Oxford, OX1 3PU, United Kingdom}

\author{Sebastian G\"ode}
\address{European XFEL, D-22869 Schenefeld, Germany}

\author{Zuzana Kon\^opkov\'a}
\address{European XFEL, D-22869 Schenefeld, Germany}

\author{Mikako Makita}
\address{European XFEL, D-22869 Schenefeld, Germany}

\author{Mikhail Mishchenko}
\address{European XFEL, D-22869 Schenefeld, Germany}

\author{Motoaki Nakatsutsumi}
\address{European XFEL, D-22869 Schenefeld, Germany}

\author{Lisa Randolph}
\address{European XFEL, D-22869 Schenefeld, Germany}

\author{Sebastian Schwalbe}
\address{Center for Advanced Systems Understanding (CASUS), D-02826 G\"orlitz, Germany}
\address{Helmholtz-Zentrum Dresden-Rossendorf, D-01328 Dresden,  Germany}

\author{Jan Vorberger}
\address{Helmholtz-Zentrum Dresden-Rossendorf, D-01328 Dresden,  Germany}


\author{Ulf Zastrau}
\address{European XFEL, D-22869 Schenefeld, Germany}

\author{Tobias Dornheim}
\address{Center for Advanced Systems Understanding (CASUS), D-02826 G\"orlitz, Germany}
\address{Helmholtz-Zentrum Dresden-Rossendorf, D-01328 Dresden,  Germany}

\author{Thomas R. Preston, D-22869 Schenefeld, any}
\ead{thomas.preston@xfel.eu}
\begin{abstract}
We report on results from an experiment at the European XFEL where we measured the x-ray Thomson scattering (XRTS) spectrum of single crystal silicon with ultrahigh resolution. Compared to similar previous experiments, we consider a more complex scattering setup, in which the scattering vector changes orientation through the crystal lattice. In doing so, we are able to observe strong geometric dependencies in the inelastic scattering spectrum of silicon at low scattering angles. Furthermore, the high quality of the experimental data allows us to benchmark state-of-the-art TDDFT calculations, and demonstrate TDDFT's ability to accurately predict these geometric dependencies. Finally, we note that this experimental data was collected at a much faster rate than another recently reported dataset using the same setup, demonstrating that ultrahigh resolution XRTS data can be collected in more general experimental scenarios.
\end{abstract}

\maketitle


\section{Introduction}\label{sec1}
The electronic dynamic structure factor (DSF) $S(\bm{q}, \omega)$ of a system is rich with information on its electronic properties. The DSF depends on a number of important system properties -- such as its temperature, density, and ionization -- and contains information on electron correlations and their localization around ions~\cite{siegfried_review,Gregori_PRE_2003, wdm_book, Dornheim_review}.
Experimentally, the DSF can be probed directly using the x-ray Thomson scattering (XRTS) technique~\cite{sheffield2010plasma}. By probing a sample with photons of incident energy $E_i$ and measuring the spectrum of the scattered photons $E_s = E_i-\hbar \omega$ along a particular scattering vector $\bm{q}$, the resulting XRTS spectrum $I(\bm{q}, E_s)$ is the convolution of the DSF with the combined source-and-instrument function (SIF) $R(E_s)$ of the experiment~\cite{siegfried_review,sheffield2010plasma,Dornheim_T2_2022}:
\begin{equation}
    I(\bm{q}, E_S) = R(\bm{q}, E_s) * S(\bm{q}, E_s-E_i) \, .
\end{equation} 
Given the potential of XRTS to provide the full information about a system and its properties, it has emerged as a leading diagnostic for studying matter in experiments, particularly in the research of matter at extreme conditions. Due to the characteristic extreme temperatures, pressures and densities of these states of matter, such conditions can only be maintained in experiments for very short times, and necessitate the use of \emph{in-situ} diagnostics that can fully probe the conditions of the sample in this short duration.
The rigorous diagnosis of matter in extreme conditions is of great interest. For one, such conditions are very common in the universe and are found in numerous astrophysical objects~\cite{Benuzzi_Mounaix_2014,becker,Kritcher2020, Haensel}. Second, a number of technological advancements in materials discovery~\cite{Kraus2016,Kraus2017,Lazicki2021, Recoules_prl_2006, nguyen2022direct} and inertial fusion energy~\cite{hu_ICF,Betti2016, Kafka_2024} utilise matter in extreme conditions, and further developments will benefit from the ability to accurately measure system properties.

X-ray Thomson scattering is therefore a potentially very powerful diagnostic, if one knows how to extract information from the scattering spectrum. In practice, however, this is very challenging. First, the SIF needs to be removed from the XRTS signal in order to extract any properties~\cite{Dornheim_T2_2022}. Owing to both the instability of deconvolution to noise and the finite spectral range of the spectrometer, direct deconvolution is not possible. Additionally, the instrument function of the spectrometer is typically non-trivial~\cite{Gawne_2024_SIF}, which adds another layer of uncertainty in whatever deconvolution approach is used.
Second, information is encoded in the shape of the DSF and directly extracting that information can, depending on the property of interest, be difficult.
For both these reasons, so-called ``forward fitting'' has become the \emph{de-facto} method for analysing XRTS data. In this approach, a model of the DSF is fit to the XRTS data, accounting for the SIF broadening, with available models varying wildly in levels of detail and complexity.

On the simplest end, there is the Chihara decomposition~\cite{Chihara_JoP_1987,Chihara_JoP_2000,Gregori_PRE_2003}, which treats systems in a chemical picture, and allows for the extremely rapid estimation of system conditions. 
While often a decent agreement between model and experiment can be achieved, this does not guarantee a correct interpretation of the measurement to the large number of free parameters in the Chihara approach; for example, B\"ohme \emph{et al.}~\cite{boehme2023evidence} have recently shown that the physically mandated free-bound contribution to the spectrum, which has been neglected in previous models, was erroneously compensated by other features.
Furthermore, such chemical models neglect all the physical structure of a system, meaning detailed examination on the electronic properties of a system are not possible.

On the other end of complexity, there are full \emph{ab initio} approaches such as time-dependent density functional theory (TDDFT)~\cite{ullrich2012time,Cazzaniga_PRB_2011, Moldabekov_prr_2023, Baczewski_prl_2016}.
In this approach, the full electronic structure is self-consistently calculated, allowing for very detailed predictions of the DSF. Indeed, TDDFT is in-principle a quasi-exact method, where approximations are introduced by use of an approximate form of the exchange-correlation functional and kernel~\cite{Byun_2020, Moldabekov_JCTC_2023,Bohme_PRL_2022, Moldabekov_JPCL_2023, Moldabekov_PPNP_2024}, neither of which are known exactly. 
The added detail in the predictions of TDDFT also comes with a massive increase in computational cost over Chihara models, but the ability to understand the electronic properties of a system in detail makes this is a worthwhile expense in many situations.

Of course, we want to be certain in TDDFT's ability to accurately predict the DSF of a system, particularly as recent TDDFT simulations predict surprising changes in the DSF of materials undergoing isochoric heating~\cite{Moldabekov_2024_Excitation, moldabekov2024ultrafast}. Direct comparison of the predictions of TDDFT with experiment seems a natural starting point, however the aforementioned use of models to interpret scattering spectra already presents a problem: theory is used to interpret experiment, which in turn is being used to benchmark the theory, etc.. Useful experimental benchmarking therefore requires many variables, such as the temperature and density of the system, to be as controlled as possible. Furthermore, the convolution with the SIF also makes direct comparisons challenging as the broadening obscures features, and can therefore allow a number of potential DSFs to produce similar looking XRTS spectra.

One approach to dealing with the SIF is to use a setup in which it is so narrow and simple that is has a negligible impact on the measured spectrum. This would then allow for the direct comparison of a predicted spectrum with experiment. Such an ultrahigh resolution setup has recently been demonstrated at the HED Scientific Instrument at the European XFEL~\cite{Gawne_2024_Ultrahigh}, providing a spectral range of 10s of eV but with a resolution $\sim$0.1~eV.
Comparisons between the scattering spectra of ambient aluminium (Al) measured at multiple scattering angles and the predicted DSFs by TDDFT were also performed, and it was shown that once the broadening effect from the finite size of the spectrometer was properly accounted for, the predictions of TDDFT matched the experimental data very well. 
This represented a promising result that even relatively simple TDDFT calculations can make accurate predictions of the electronic response of simple metals.

As a material, Al is quite a simple system to model: Al foils are polycrystalline, so any measured signal would be the average over all lattice orientations. In any case, the TDDFT-predicted DSF was found to be relatively isotropic with respect to the direction of the scattering vector. In other words, simply calculating the Al DSF with a scattering vector along the [100] direction was sufficient to model the spectrum.
Moreover, Al is a simple metal, and so the treatment of exchange and correlation on the level of a free electron gas via the (adiabatic) local density approximation was expected to be quite accurate.
Silicon (Si), on the other hand, is a much more complex material. First, it is a covalently bonded semi-conductor, which pushes the limits of applicability of approximate exchange-correlation functionals. Second, wafers of Si can be grown as monocrystals with a particular orientation, meaning any dependencies of the electronic response of the material on orientation can be examined. Indeed, this is what has been observed in previous measurements of the bulk Si plasmon in electron energy loss spectroscopy (EELS) experiments~\cite{Stiebling_1978_Dispersion,Chen_1980_Bulk} and inelastic x-ray scattering spectroscopy measurements at synchrotrons~\cite{Schuelke_1995_Dynamic,Weissker_2010_Dynamic}. The plasmon dispersion in Si is different along the [100], [110], and [111] directions~\cite{Stiebling_1978_Dispersion,Chen_1980_Bulk}, and the shapes of the scattering spectra are also quite different along these directions~\cite{Schuelke_1995_Dynamic,Weissker_2010_Dynamic}. For accurate simulations of Si, this geometry-dependence should therefore be important to consider, and add a new level of complexity to modelling the scattering spectra with TDDFT. Indeed, some simulation results suggest TDDFT requires modifications in order to produce the correct scattering spectrum~\cite{Weissker_2010_Dynamic}.

Here we report on ultrahigh resolution measurements of the bulk plasmon in a  single crystal of Si (100), using the x-ray free electron laser (XFEL) at the European XFEL in Germany. Inelastic scattering spectra are collected over a range of scattering angles between $3.6^\circ$--$25.6^\circ$, with a resolution $\sim0.1$~eV, and a spectral range of up to $\sim 70$~eV.
With a novel scattering setup, we demonstrate TDDFT's ability to accurately simulate complex geometric effects in a monocrystalline semi-conductor system.
As with Al, we find accounting for the finite size of the spectrometer is a necessary consideration for the accurate modeling of the experimental data at low scattering vectors, without the need for energy-dependent broadening, in contrast to Ref.~\cite{Weissker_2010_Dynamic}.

Lastly, we report on two changes of the setup over the one reported in Ref.~\cite{Gawne_2024_Ultrahigh}. First, the spectral window in this dataset reaches up to $70$~eV below the elastic, which is much larger than the $40$~eV window used previously.
Second, the experimental data was collected at a much faster rate than in Ref.~\cite{Gawne_2024_Ultrahigh}, but the signal-to-noise ratio is still sufficient to perform benchmarking of TDDFT simulations. We conclude then that this offers a promising outlook for performing ultrahigh resolution measurements in more generic experimental scenarios.

\section{Experiment}\label{s:exp}

Measurements of the inelastic scattering of silicon were performed at the HED instrument at the European XFEL in Germany~\cite{Zastrau2021}. The setup used is identical to that of recently reported measurements of the Al plasmon~\cite{Gawne_2024_Ultrahigh} and took place during the same beamtime. A more detailed explanation of the setup is provided in Ref.~\cite{Gawne_2024_Ultrahigh}, but we will also summarise the important elements here and the highlight the differences in approach.

First, the targets used here here were 50~$\mu$m thick monocrystalline Si (100) wafers. The Si has a diamond cubic crystal structure, which means it consists of two intersecting face-centered cubic (fcc) lattices ($\it{Fd\overline{3}m}$ with an additional atom at $(1/4, 1/4, 1/4)a$, where $a$ is the lattice constant in the conventional unit cell). As a single crystal, the wafers have a well-defined normal along the (100) direction, and this normal was maintained in fixed alignment with the incoming XFEL beam direction.

The XFEL beam was self-seeded to an energy $E_0 \sim 7703$~eV, and then passed through a four-bounce Si (111) monochromator with an acceptance range of 0.8~eV to remove the underlying self-amplified spontaneous emission (SASE) pedestal.
The beam is then focused onto the target to a spot size of $\sim 10$~$\mu$m.
The scattered x-rays were collected on a spherically-bent Si (533) diced crystal analyser (DCA)~\cite{Descamps_SR_2020,Wollenweber_RSI_2021}. This DCA has recently been demonstrated as being capable of measuring x-ray scattering spectra with an energy resolution $\sim 0.1$~eV and spectral range of several 10s of eV~\cite{Gawne_2024_Ultrahigh}.
The scattered x-rays are finally recorded on a Jungfrau detector~\cite{Mozzanica2018} with asymmetric pixels -- 25~$\mu$m long in the dispersive direction and 225~$\mu$m in the non-dispersive direction -- in order to minimise pixel broadening in the dispersive direction. From the spectrometer calibration, the energy dispersion was determined to be $22.54\pm0.15$~meV/pixel~\cite{Gawne_2024_Ultrahigh}.

Unfortunately, the quality of the self-seeded beam was unsatisfactory during the beam time, and $\sim85$~\% of the beam fluence was in the SASE pedestal. Therefore, after the beam was passed through the monochromator and transmitted through the beamline optics, only 15.5--21.8~$\mu$J of energy was measured on target by a gas monitor before the target~\cite{Gawne_2024_Ultrahigh}, which means very few photons are available to be scattered. However, this also meant that the intensity on target was too low to heat the samples, which simplifies the modeling as both the temperature and density can be taken to be ambient.

The DCA has a spectral window of 3.5~eV, but the electronic response energy scale is over 10s of eV. Therefore, to measure the spectrum across a wider spectral range, the spectral window is moved by rotating the DCA to change the Bragg angle. The DCA is then held at each position for a fixed number of frames, collected at a rate of 10~Hz, before shifting the spectral window again. Each frame is integrated over 20 x-ray pulses, with the pulses having a repetition rate of 2.2~MHz.
It is necessary to hold the DCA in a given position for a period of time in order to collect enough photons to measure the spectrum to the desired signal-to-noise ratio (SNR).
The full spectrum is then stitched together from these different windows.
In the case of the previous Al measurements~\cite{Gawne_2024_Ultrahigh}, the spectral window was held in each position for 20--40~s (200--400 frames). This resulted in very high quality spectra with low signal to noise, but meant that the collection times were very long in order to build a full spectrum (this is partly due to the poor beam quality, meaning a low number of photons were incident on the target to then scatter). The spectral range was also limited to 40~eV, as beyond this point no scattering was expected.
For the present Si data, the ability of the DCA to act as an efficient spectrometer was tested. To do this, the DCA was held in each spectral window for only 5~s (50 frames), which results in a much faster collection time. Coincidentally, Si also has a weaker inelastic signal than Al at the same scattering vectors, which is predicted by both TDDFT and is what is observed in experiment by roughly by the same amounts (up to $\sim5.5$ times weaker at the peak). Nevertheless, despite the shorter collection times and the weaker scattering signal, the quality of the spectra is still very high and more than sufficient quality for benchmarking TDDFT calculations.

\begin{figure}
    \centering
    \includegraphics[width=\textwidth,keepaspectratio]{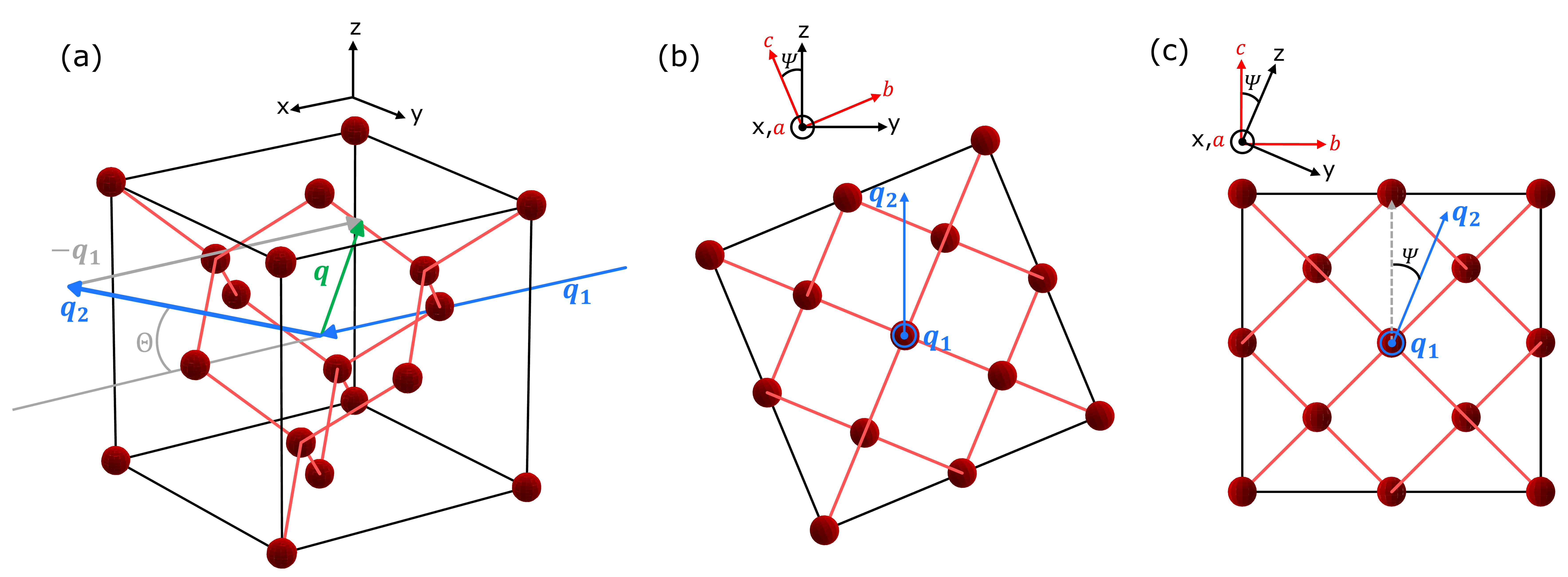}
    \caption{
    (a) A schematic of the experimental geometry with respect to the crystal lattice.
    The red spheres represent the Si atoms. The black cube wireframe represents the conventional unit cell, while the red lines connecting the spheres inside the cell represent the Si bonds connecting the nearest-neighbour atoms.    
    The beam direction $\bm{q_1}$ and the normal of the Si (100) crystal are both aligned the $(1,0,0)$ direction. The scattered rays travel along a vector $\hat{\bm{q_2}} = (\cos\Theta, 0, \sin\Theta)$, where $\Theta$ is the scattering angle. The spectrum for a given scattering vector $\bm{q} \equiv \bm{q_2} - \bm{q_1}$ is measured by moving the center of the DCA the chosen scattering angle.
    (b) The crystal lattice can be rotated about the beam by an angle $\psi$ without affecting the alignment of the crystal normal and the beam direction, but it changes the alignment of $\bm{q_2}$ and $\bm{q}$ through the crystal lattice.
    (c) Equivalent to rotating the lattice about the beam direction, $\bm{q_2}$ can be rotated about the beam direction by the angle $-\psi$ -- this approach is used to define $\bm{q}$ in the TDDFT calculations.
    }
    \label{fig:Setup}
\end{figure}

To measure the scattering spectrum at different scattering angles, the entire setup is rotated so that the DCA can capture and reflect the light travelling along a scattering angle of $\Theta$ (see Fig.~\ref{fig:Setup}~(a)) to the detector. In total, the scattering spectra at five central scattering angles between $3.6^\circ$--$25.6^\circ$ were measured. As the DCA has a finite size, it collects photons over a range of scattering vectors resulting in a spectrum that is broadened by so-called $q$-vector blurring. To limit this broadening, an Al slit mask was placed on the DCA which reduces the angular coverage to $\pm 1.4^\circ$, or a $q$ coverage of $\pm0.095$~$\textup{\AA}^{-1}$ to $\pm0.098$~$\textup{\AA}^{-1}$. The consideration of this effect is discussed in the remaining sections of the manuscript as it is very important for correctly interpreting results~\cite{Gawne_2024_Ultrahigh}.

As with the Al data in Ref.~\cite{Gawne_2024_Ultrahigh}, the data is smoothed to tease out the signal from the noise. Here, however, we use a moving average over a five pixel window in order to get an estimate of the spectral uncertainty from the variance to this average. This is important as the uncertainty in the spectral shape is now dominated by noise rather than the calibration uncertainty.

%

\subsection{Experimental Geometry}
Due to the importance of the experimental geometry on the results, we now consider this in some detail. A schematic of the experimental geometry through the conventional unit cell of Si is shown in Fig.~\ref{fig:Setup}~(a). The Si sample was a single crystal with its normal oriented along the [100] direction. This normal was aligned along the direction (wave vector) of the incident beam $\bm{q_1} = Q(1, 0, 0)$ for all measurements, where $Q = 2\pi E/hc$, $E$ is the photon energy of the beam, $h$ is Planck's constant, and $c$ is the speed of light. The center of the DCA is then positioned to measure photons emerging along the wave vector $\bm{q_2} = Q_2(\cos\Theta, 0, \sin\Theta)$, where $\Theta$ is the (central) scattering angle. Note that we consider the energy loss to be small so that $Q_2\approx Q$.
The (central) scattering vector of the measured spectrum is $\bm{q} \equiv \bm{q_2} - \bm{q_1}$. Clearly, this is not oriented along a fixed direction in the Si lattice, but instead varies in orientation. This is a crucial difference to other previous experiments on Si~\cite{Stiebling_1978_Dispersion,Chen_1980_Bulk,Schuelke_1995_Dynamic,Weissker_2010_Dynamic} where the sample is tilted such that the normal of the lattice and the scattering vector remain aligned, which allows for the DSF to be studied along this fixed direction. In our setup, we target more complex geometric behaviours of the DSF as the scattering vector can be aligned anywhere through the crystal.

We note that the orientation of $\bm{q}$ through the crystal lattice is not fully defined by the orientation of the crystal lattice normal along the beam direction as it is possible to rotate the lattice about the beam direction without affecting the direction of the normal, as is shown in Fig.~\ref{fig:Setup}~(b). In other words, if $\psi$ is the angle by which the lattice is rotated about the beam direction (and crystal normal), we can define the lattice vectors of the conventional unit cell in the experimental coordinates as:
\begin{equation}
    \bm{a} = (1, \, 0, \, 0)\, , \,\,\,\,\,\,\,\,
    \bm{b} = (0, \, \cos\psi, \, \sin\psi)\, , \,\,\,\,\,\,\,\,
    \bm{c} = (0, \, -\sin\psi, \, \cos\psi)\, 
\end{equation}
For our simulations, it is somewhat inconvenient to change the definition of the lattice parameters as the atom positions are also defined in real space. Instead, as seen in Fig.~\ref{fig:Setup}~(b) and~(c), the equivalent direction of $\bm{q}$ through the crystal lattice can be retrieved by rotating $\bm{q_2}$ about the beam direction by an angle $-\psi$. We therefore define the set of wave vectors as:
\begin{equation}
    \bm{q_1} = Q(1, \, 0, \, 0)\, , \,\,\,\,\,\,\,\,
    \bm{q_2} = Q(\cos\Theta, \, \sin\psi \sin\Theta, \, \cos\psi \sin\Theta)\, , \,\,\,\,\,\,\,\,
    \bm{q} = \bm{q_2} - \bm{q_1} \, .
    \label{eq:Qs}
\end{equation}
This yields an ambiguity in the actual direction the scattering vector passes through the crystal lattice. Unfortunately, this angle $\psi$ was not measured during the experiment as, although an area detector was present to measure diffraction, we were unable to rotate the sample sufficiently to measure the diffraction spots.
However, as will be shown, because Si is so sensitive to the direction of the scattering vector through the lattice, this makes it possible to infer what this orientation was from the TDDFT calculations.

\begin{figure}
    \centering
    \begin{subfigure}
        \centering
        \includegraphics[width=0.45\textwidth,keepaspectratio]{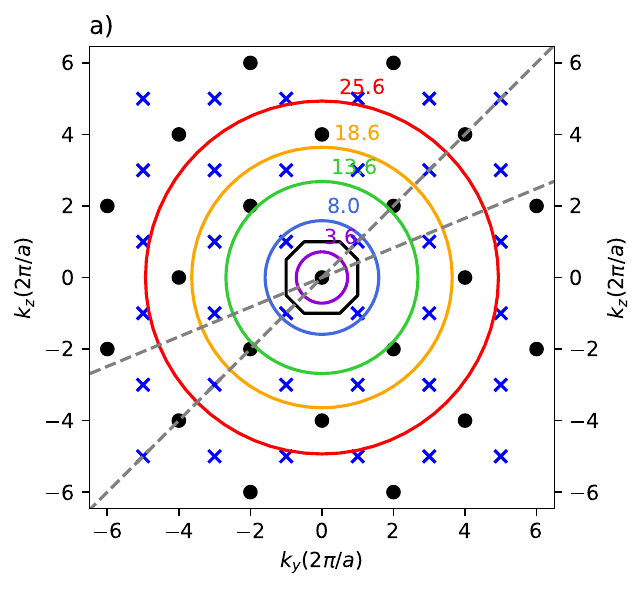}
    \end{subfigure}%
    ~ 
    \begin{subfigure}
        \centering
        \includegraphics[width=0.45\textwidth,keepaspectratio]{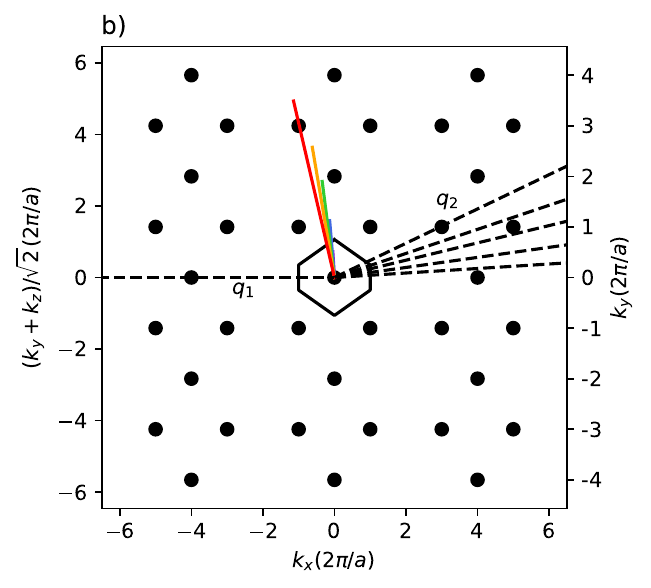}
    \end{subfigure}%
    ~ 
    \caption{The Si lattice in reciprocal space in the conventional unit cell. Shown in a) is the $k_x = 0$ plane in units of $2\pi/a$ where $a=5.4309\textup{\AA}$ is the lattice constant for the conventional unit cell in Si. The incoming photon $q_{1}$ is aligned with the [100] direction (out of the page) and scatters with various $q$ depending on the observed $q_{2}$. The orientation around the [100] direction is unknown, therefore the $q$ are shown as circles, labelled by $\Theta$ scattering angle. The reciprocal lattice points with Miller indices $[hkl]$, found in the plane $[0kl]$ are shown in black, and for completeness reciprocal lattice points in the $k_x = -1$ plane are shown as blue crosses, i.e. $[\overline{1}kl]$. Also indicated are the planes at $45^\circ$ and $22.5^\circ$ with grey dashed lines. In b) the $k_y=k_z$ plane at $45^\circ$ is shown. Here it is clear that as the scattering angle increases the probed $q$ tilts back. The reciprocal lattice points found in the plane $[hkk]$ are shown in black. 
    In both, the first Brillouin zone of the fcc lattice is indicated in black around the central $\Gamma=[000]$ point, forming a truncated octahedron~\cite{ashcroft1976solid}.
    }
    \label{fig:Si_geom}
\end{figure}

Changing the scattering angle changes both the length of the scattering vector and, in the present work, its direction through the crystal lattice. As the target is a single crystal, this effect is best illustrated by considering the scattering geometry in reciprocal space~\cite{ashcroft1976solid}.
In Fig.~\ref{fig:Si_geom}~(a), the reciprocal lattice points of the Si lattice in the $k_x=0$ plane.
When a scattering vector $\bm{q}$ lies on one of the points, the Laue vector equation is satisfied, and the relationship between the scattering angle $\Theta \equiv 2\theta$, the wavelength of the incident radiation $\lambda$ and the separation of a lattice planes $d_{hkl}$ for Miller indices $[hkl]$ is given by the familiar Bragg equation:
\begin{equation}
    \lambda =2d_{hkl} \sin(\theta) \, .
\end{equation}
The possible orientations of the different vectors for each scattering angle $2\theta$ are represented as circles, with $\psi$ determining where the vector is pointing to on the circle. The length of a scattering vector is given by $q=2Q \sin(\theta)$, so higher angles correspond to probing longer distances in reciprocal space, and therefore shorter length scales in real space.

A key observation is that at the smallest scattering angle of $3.6^\circ$, the scattering vector does not extend beyond the first Brillouin zone (BZ) -- indicated by the solid black octagon -- and the bulk behaviour of the system is being probed. In other words, the scattered photon is mostly oblivious to the specifics of crystal structure as the response of the system over multiple unit cells is probed, and the DSF should therefore be largely insensitive to the specific orientation of the scattering vector.
At the second smallest scattering angle of $8.0^\circ$, the scattering vector will be slightly beyond the surface of the first Brillouin zone and should only just begin to probe the structural dependencies of the Si lattice.
At higher scattering angles the length of $\bm{q}$ in reciprocal space is now sufficient to extend beyond the first BZ, so the orientation of $\bm{q}$ relative to the lattice planes should now contribute significantly to the DSF.
Furthermore, increasing the scattering angle in this setup increases the tilt of the scattering vector through reciprocal lattice space, which makes different families of points available, as shown in Fig.~\ref{fig:Si_geom}.
Depending on the specific angle $\psi$, the lattice vectors can pass various nearby lattice points, or could even lie on them to produce a diffraction spot. Unlike diffraction peaks, which require the scattering vector to lie on a specific reciprocal lattice point~\cite{ashcroft1976solid}, in XRTS the scattering vector can be positioned anywhere to probe the electronic response in that particular direction. Nevertheless, the presence of these reciprocal points will still affect the electronic response as the electronic structure of the system looks different along different lattice planes. And, as the scattering vector is changing, it passes non-trivially in the vicinity of different points. We therefore anticipate strong geometric effects on the DSF will be exposed in this experiment.

For completeness, we note that at a sufficiently high scattering angle (which is not reached in this experiment) the electronic response should become increasingly agnostic to the crystal structural again as the probing length scale becomes shorter than the separation of the closest lattice planes in real space. And at very high angles, the system is eventually being probed on the scale of individual particles and so the orientation of the scattering vector should become even less important.

%


\section{XRTS spectrum of silicon in a semiconducting crystal diamond state}\label{s:results}
\begin{figure}
    \centering
    \includegraphics[width=0.7\textwidth,keepaspectratio]{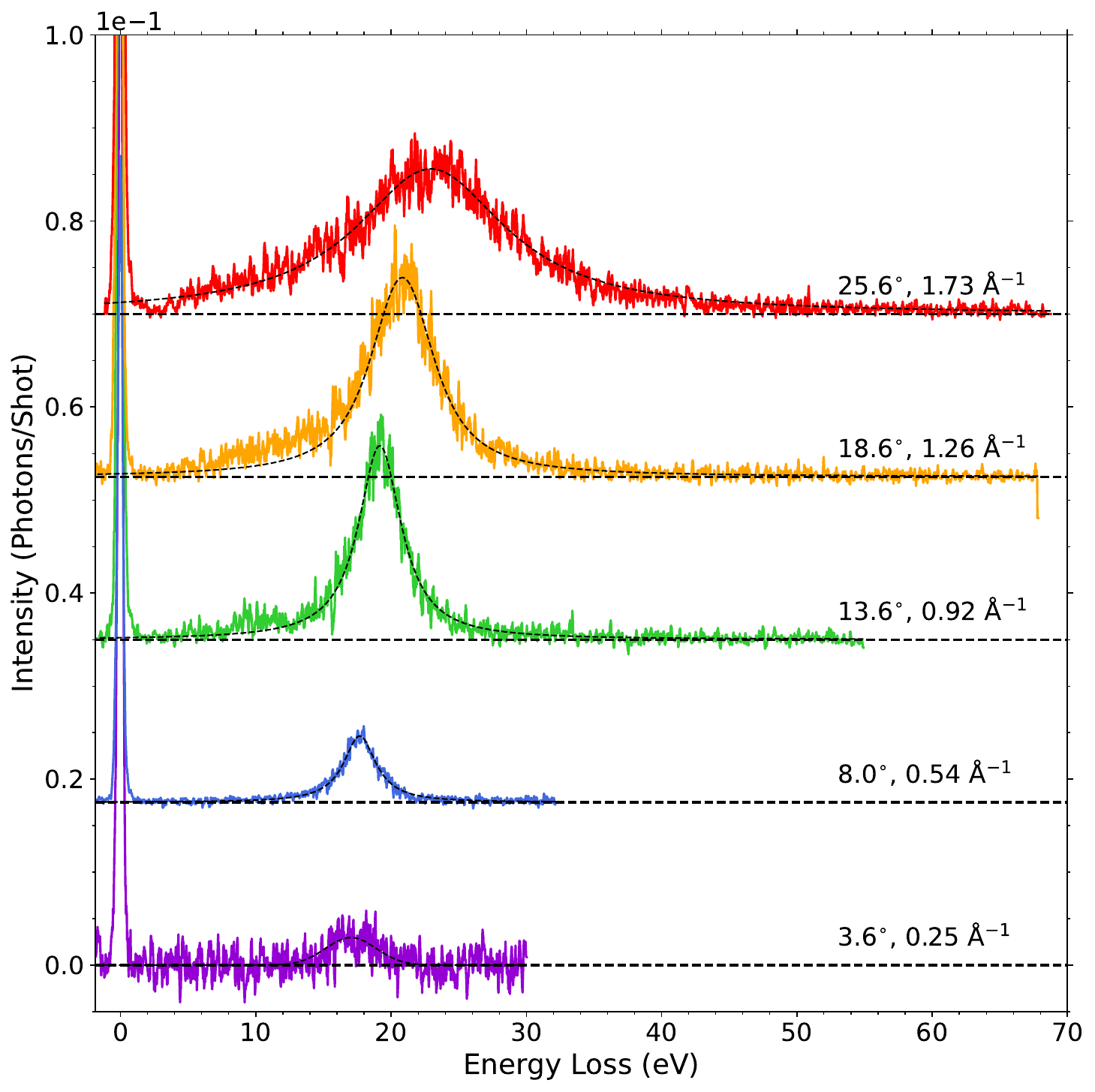}
    \caption{Measured XRTS intensity for five different wavenumbers as a function of the photon energy loss $E=E_0-E_s$ in units of integrated intensity in photons/shot. The curves are offset vertically for clarity and show the variation in position, intensity, and shape of the plasmon in silicon. The dashed lines over each curve shows the Voigt profile fitted to each peak which is used to determine the maximum position.
    }
    \label{fig:Plasmon}
\end{figure}

The measured inelastic scattering spectrum of Si for the different scattering vectors are plotted in Fig.~\ref{fig:Plasmon}. For each spectrum, a clear peak is visible, which changes in intensity, width, and position as the scattering angle is varied. 
For the four highest scattering angles, the SNRs of these spectra are more than sufficient to observe the specific shapes of the inelastic spectra. The lower noise of the spectrum at $8^\circ$ is because the collection time per spectral window was 30~s rather than the 5~s used for the other scattering angles. This spectrum is comparable in quality to the Al spectra in Ref.~\cite{Gawne_2024_Ultrahigh} and demonstrates the improvement in the SNR by using longer collection times, or increasing the beam energy in each pulse, to collect more scattered photons. However, for the highest three scattering angles, this quality is achieved with a collection rate that was 4--6 times faster than that of the equivalent Al data~\cite{Gawne_2024_Ultrahigh}, on a material that scatters more weakly than Al.

At the lowest scattering angle, the spectrum is significantly more noisy than the remaining spectra because the inelastic scattering intensity scales as $\sim q^2$, so simply very few photons are scattered inelastically at this angle. Nevertheless, a peak is still clearly visible at an appropriate position, based on the other scattering angles.
However, actually identifying the maximum of the peak proved challenging as when the DCA was scanning around the peak of the scattering, the FEL beam energy dropped quite substantially, resulting in the slight dip in the intensity around 17~eV. Even though the beam energy is accounted for in the normalisation of the spectra, the lack of photons in this region nevertheless makes it hard to identify the true peak position of the scattering. Unfortunately, due to time constraints, it was not possible to collect further data at this scattering angle, and the overall noise level makes it challenging to compare the shape of this peak to TDDFT. Still, it is clear that this very weak inelastic scattering feature was nevertheless detected, and longer collection times would be able to resolve this feature in more detail. This suggests that one approach to using this setup would be to use a quick scan with the DCA to identify the position of features, then do a more focused scan at the position of features to achieve the desired level of signal-to-noise.

\begin{figure}
    \centering
    \includegraphics[width=0.7\textwidth,keepaspectratio]{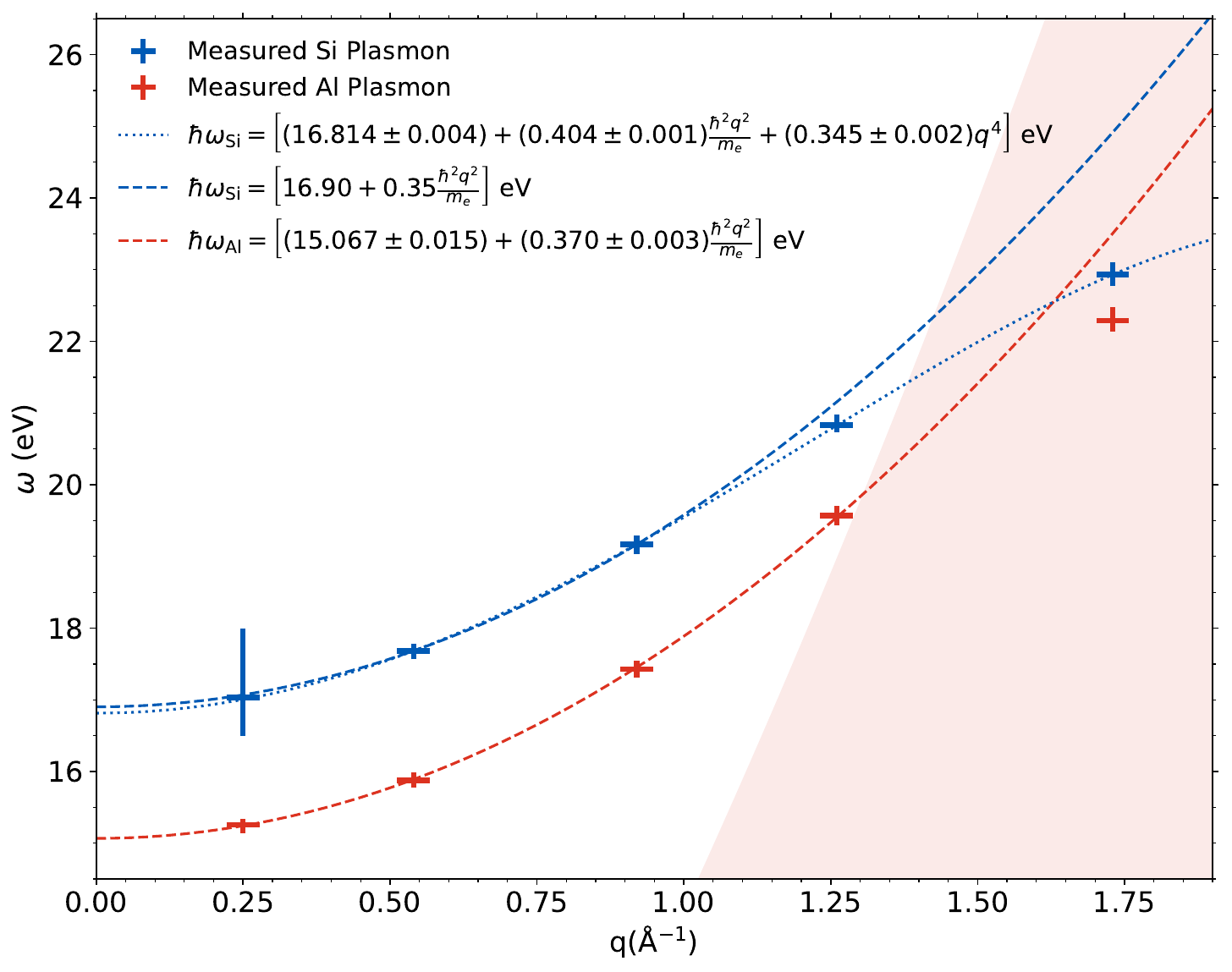}
    \caption{Dispersion of the Si (blue) and Al plasmon (red) from Ref.~\cite{Gawne_2024_Ultrahigh}, including fits to these points (dashed lines) using the Bohm-Gross relation (Eq.~(\ref{eq:BohmGross})), and a second quartic fit for Si (dotted). Uncertainties in the fit parameters account for the calibration uncertainty and the $q$-vector blurring. The red shaded area indicates the pair-continuum region for Al. The large uncertainty bar on the lowest Si plasmon is due to a gap appearing in the peak of the spectrum, making it challenging to unambiguously identify the maximum position of the peak, and it is not used in the fits for Si, hence no uncertainty can be calculated on the the Si Bohm-Gross fit. While Si is expected to obey the Bohm-Gross relationship for $q\lesssim1.2 \, \textup{\AA}^{-1}$, this is only when the scattering is along fixed orientation, which in this experiment is changing. These fit parameters for Si therefore differ substantially from the literature.}
    \label{fig:Dispersion}
\end{figure}

In order to examine the dispersive behaviour of the Si spectra with respect to changing wave vector, we identify the maxima of these peaks by fitting Voigt profiles (\textsc{scipy.special.voigt\_profile} from the SciPy package for Python~\cite{2020SciPy-NMeth}) to the peaks and taking the maxima. Clearly the shapes of the peaks do not have the symmetric shape that is characteristic of a Voigt profile, however around the maxima this approximation appeared to be sufficiently good to extract the position of the maxima.
These maxima are plotted in Fig.~\ref{fig:Dispersion}, along with the equivalent Al plasmon points~\cite{Gawne_2024_Ultrahigh} for reference. The vertical uncertainty bars come from the calibration uncertainty (described in the Supplementary Material of Ref.~\cite{Gawne_2024_Ultrahigh}), except for the lowest scattering vector in Si. For this point, the uncertainty is dominated by the difficulty in identifying the position of the maximum from the relatively flat plateau: the center of the cross indicates the position from the fit at the nominal calibration, while the limits of the vertical bar shows the approximate width of the flat region around the peak.

The horizontal bars of each point require more careful interpretation: they do not represent the uncertainty in the central scattering vector, as the center of the DCA was initially positioned carefully at a scattering angle of $13.6^\circ$, and its position (and so central scattering vector) was then carefully controlled with motors that have high precision. Instead, as already explained, the DCA covers a finite angular range and so collects spectra over a range of scattering vectors of $\sim \pm 0.1 \, \textup{\AA}^{-1}$. As the number of die in the DCA is uniform in $q$, it reflects the signal from this range of scattering vectors uniformly in each energy bin. Therefore, we plot the standard deviation of a uniform distribution in Fig.~\ref{fig:Dispersion} as the horizontal bars, giving an accuracy of $\sim \pm 0.03 \, \textup{\AA}^{-1}$.

For a free electron gas, the plasmon dispersion is expected to follow a quadratic from the Bohm-Gross relation~\cite{Bohm_Gross}
\begin{equation}
    \hbar\omega(q) = \hbar\omega_p + \alpha \frac{\hbar^2 q^2}{m_e} \, ,
    \label{eq:BohmGross}
\end{equation}
where $\omega_p$ is the plasma frequency, and $\alpha$ is a scaling prefactor. 
The measured Al plasmon dispersion follows this quadratic essentially exactly for the four points below its electron--hole pair continuum region, and even accounting for the calibration uncertainty and $\bm{q}$-vector blurring still resulted in very tight constraints on $\omega_p$ and $\alpha$~\cite{Gawne_2024_Ultrahigh}. For the point in the pair continuum~\cite{quantum_theory}, the plasmon is damped as it decays into multiple excitations due to Landau damping, and so it no longer follows a quadratic dispersion.

Al is often considered a prototypical ``free electron gas metal'', and so it is expected that it would obey the Bohm-Gross relationship.
Although Si is a covalently-bonded semi-conductor, the photon energies here greatly exceed the band gap, so the electrons are highly mobile and behave as nearly-free electrons.
Up to a critical scattering vector $q_c \sim 1.2 \, \textup{\AA}^{-1}$, quadratic dispersion has been observed in Si along the [100], [110] and [111] directions~\cite{Stiebling_1978_Dispersion,Chen_1980_Bulk}.

In Ref.~\cite{Stiebling_1978_Dispersion}, it is noted that the plasma frequency of Si at $\hbar\omega_p = 16.6$~eV is substantially larger than the energy gap of the bound electrons ($\sim 1$~eV), allowing for this nearly-free electron treatment.
The value of $q_c$ is comparable to the length of the Si reciprocal lattice vector $2\pi/a \sim 1.16 \, \textup{\AA}^{-1}$, so along as the scattering vector is contained inside the first BZ, then nearly-free electron like behaviour can be observed. Indeed, near the bottom of the bands around the $\Gamma$-point, the shape of the bands is parabolic (the same for a free electron), but this stops being the case as the surface of the BZ is approached~\cite{Si_Band_Structure}.
Furthermore, $\omega_p$ is the same for these three orientations; i.e. the entire structure is probed simultaneously it appears isotropic (see Fig.~\ref{fig:Si_geom}~(a)).
Still, even within a scattering vector contained in the first BZ, the influence of the crystal lattice is still felt. Namely, the scaling factor $\alpha$ is different for each of the three principal directions in Si~\cite{Stiebling_1978_Dispersion,Chen_1980_Bulk}. This may be understood from the fact that the distance to the first BZ is different along each of these directions, so the system will still appear non-isotropic.

For the Si data presented here, we have only two reliable points at $q<q_c$, which is not sufficient to determine the fit to the Eq.~(\ref{eq:BohmGross}) with uncertainties. Nonetheless, fitting to the two points (plotted in Fig.~\ref{fig:Dispersion}) gives an $\hbar\omega_p = 16.9$~eV, which is substantially higher than the $\hbar\omega_p = 16.6$~eV reported in EELS measurements~\cite{Stiebling_1978_Dispersion,Chen_1980_Bulk}.
A second fit of the four highest scattering vectors to a quartic function, shown in Fig.~\ref{fig:Dispersion}, is predominately intended to guide the eye to the behaviour of the dispersion. But, this fit still implied a plasma frequency of $\hbar\omega_p \sim 16.8$~eV, which is closer but is still substantially different from the measured value~\cite{Stiebling_1978_Dispersion,Chen_1980_Bulk}.
However, this apparent discrepancies may be explained by the fact that the scaling parameter $\alpha$ itself depends on the orientation of the scattering vector through the crystal, and within the present experiment this orientation is changing.
Therefore, it is not possible to observe the quadratic behaviour here as $\alpha$ is now an unknown function of the scattering vector, and this would need to be accounted for in the fitting. Otherwise, if one wishes to make observations of the dispersive behaviour of the Si plasmon, it seems necessary to take care that the scattering vector points along a fixed direction. Given this was not a primary objective of this experiment, we are content to state that we observe dispersion, but the strong geometry dependence of the scattering makes this dispersion non-trivial.

%


\section{Predictions of the DSF using TDDFT}\label{s:simulation}

While the dispersive behaviour of the plasmons is difficult to quantify from the present experimental data, this was not the main objective of the experiment, which was to measure their shape. Still, the complexity of the dispersion is indicative that the DSF of Si will be strongly geometry dependent, and the purpose of this experiment was to collect very high quality data that could be used to benchmark theory.
In order to model the electronic response of systems, (linear response) time-dependent density functional theory has become one of the leading approaches as it is able to self-consistently capture the electronic response to a perturbation in an electronic and ionic environment~\cite{ullrich2012time,Cazzaniga_PRB_2011,Moldabekov_jcp_2023}. 

While TDDFT is formally exact, in practice it still has input approximations as the exact exchange-correlation functional and kernel are unknown, and both need some level of approximation. Given the powerful predictive potential of TDDFT, it is pertinent to examine whether the predictions it makes are indeed accurate.

Already, it has been shown that the simplest level of TDDFT, the adiabatic local density approximation (ALDA), was sufficient to accurately model DSF of Al~\cite{Gawne_2024_Ultrahigh}. However, as previously mentioned, the relative simplicity of Al meant that accounting for geometric effects was not crucial to make accurate predictions. Based on previous experiments and simulations~\cite{Stiebling_1978_Dispersion,Chen_1980_Bulk,Schuelke_1995_Dynamic,Weissker_2010_Dynamic} and the results of the present work, we expect that a good theoretical model must be able to observe strong geometry dependencies in the DSF when they are present like here. We now investigate this predictive capability of TDDFT.


\subsection{Simulation details}
The LR-TDDFT calculations for Si with a crystal diamond structure were performed using Quantum ESPRESSO \cite{Giannozzi_2009, Giannozzi_2017, Giannozzi_jcp_2020, Carnimeo_JCTC_2023, TIMROV2015460}. We used a $20\times20\times20$ $k$-point grid and an energy cutoff of $16~{\rm Ry}$.  A cubic simulation cell with a side length of $5.431~\textup{\AA}$ and a lattice parameter of the crystal $a=5.431 ~\textup{\AA}$ was considered. The results were computed using the Lorentzian smearing parameter  $\eta=0.1~{\rm eV}$. The used pseudopotential {Si.pz-vbc.UPF} is from the Quantum ESPRESSO pseudopotential database~\cite{PP_ESPRESSO_2022, Timrov_prb_2013}.
For each scattering angle, five LRTDDFT calculations were conducted within the $q$-vector blurring range. For simulations, atomic structure and the components of the scattering wavevectors with respect to the simulation cell were defined using the Atomic Simulation Environment \cite{Hjorth_Larsen_2017}.


\subsection{Geometric dependencies of the DSF in silicon}

\begin{figure}
    \centering
    \begin{subfigure}
        \centering
        \includegraphics[width=0.45\textwidth,keepaspectratio]{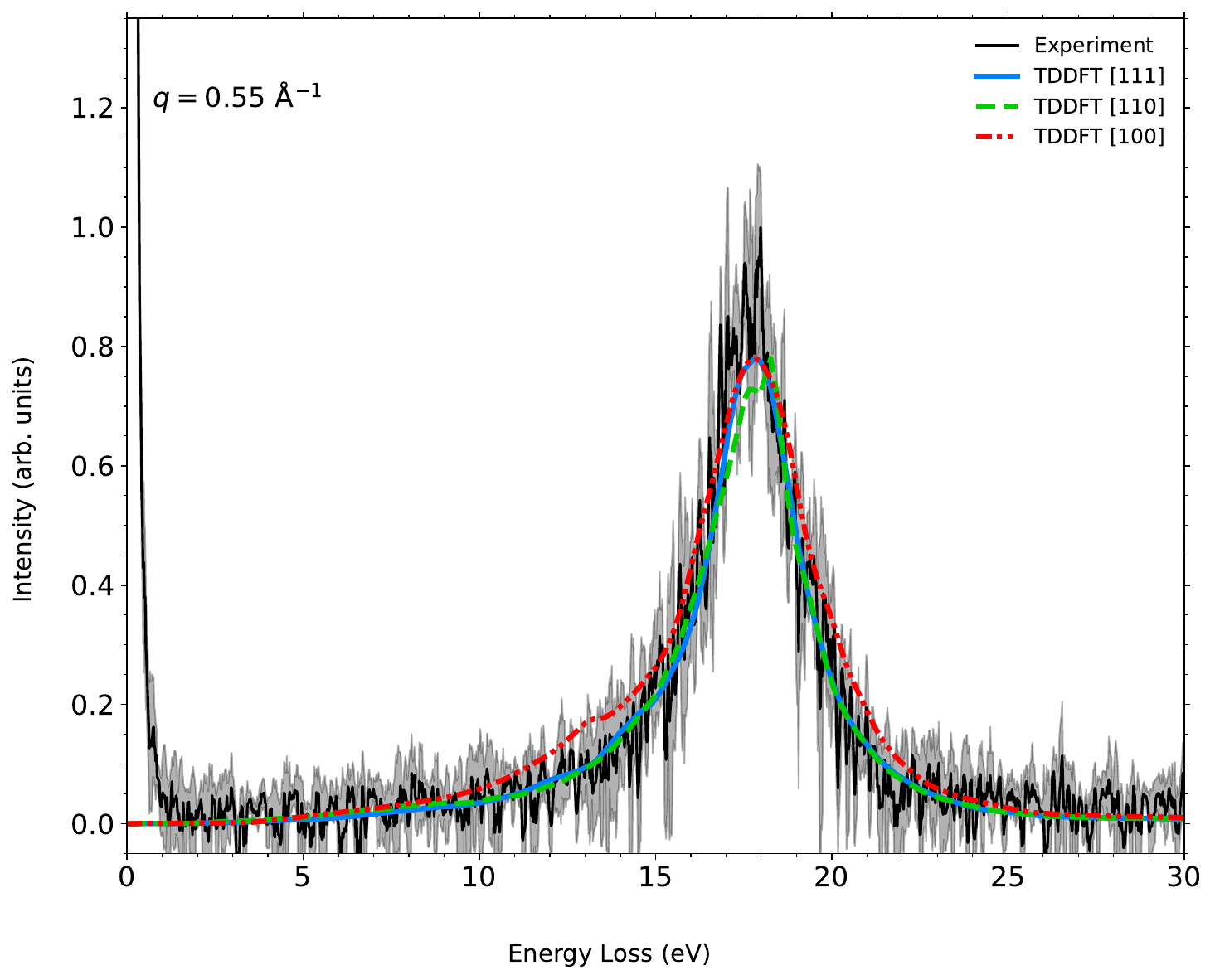}
    \end{subfigure}
    ~
    \begin{subfigure}
        \centering
        \includegraphics[width=0.45\textwidth,keepaspectratio]{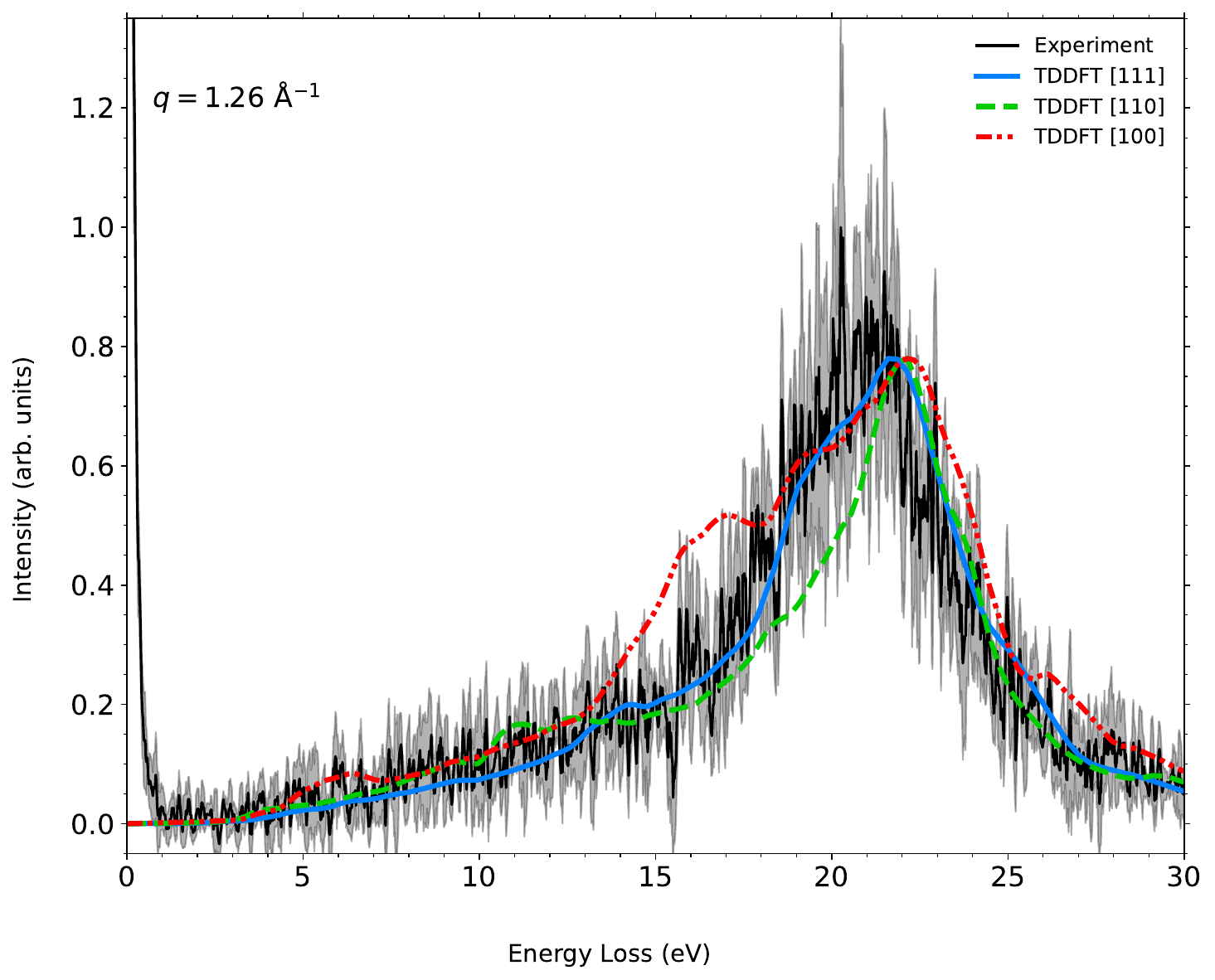}
    \end{subfigure}
    \caption{Comparison of the experimental data (black) to the TDDFT calculations of the DSF of Si along the [111] (blue solid), [110] (green dashed), and [100] (red dotted) directions, for a scattering vectors $q=0.55\textup{\AA}^{-1}$ and $1.26\textup{\AA}^{-1}$.
    }
    \label{fig:TDDFT_Principal}
\end{figure} 
\begin{figure}
    \centering
    \begin{subfigure}
        \centering
        \includegraphics[width=0.45\textwidth,keepaspectratio]{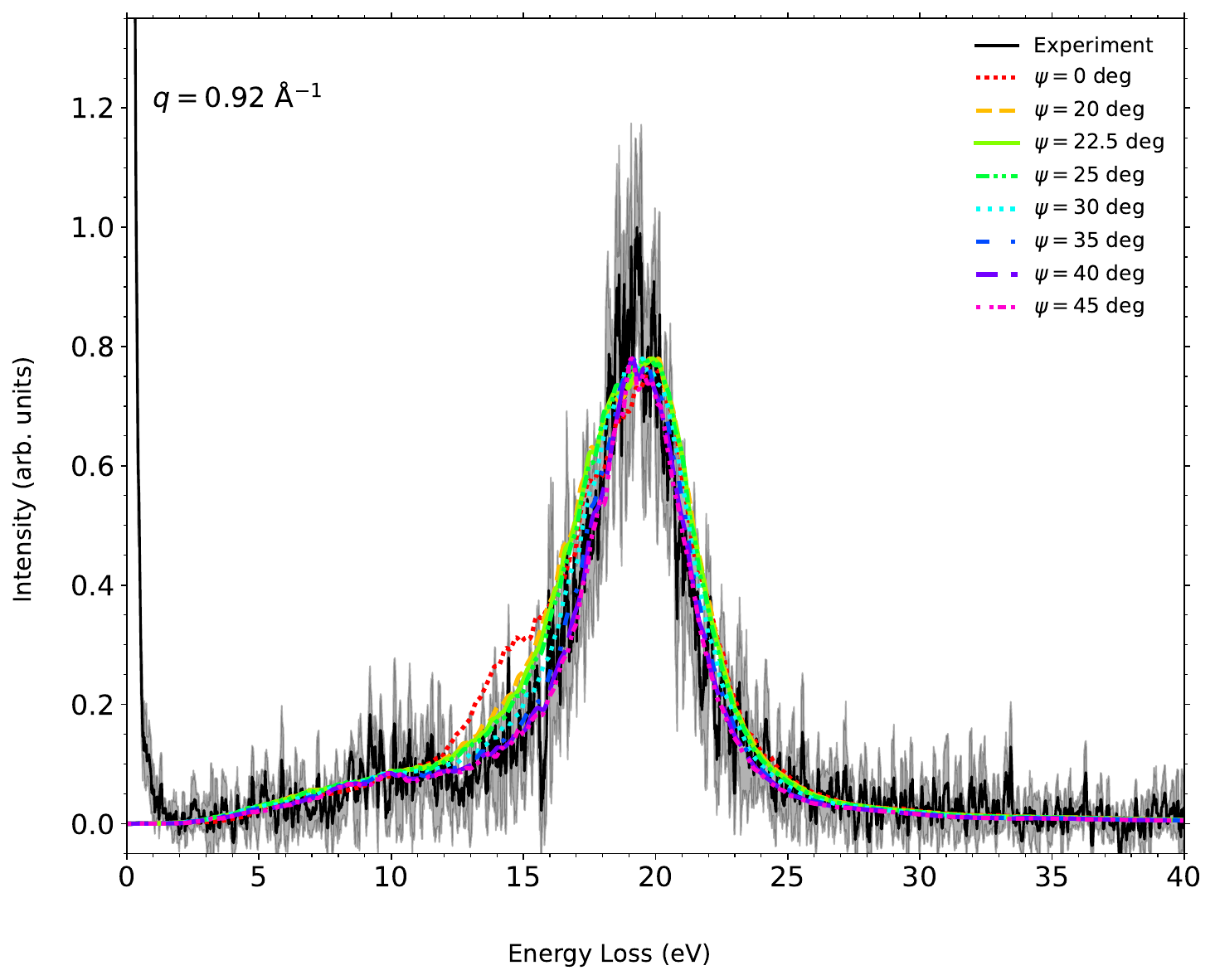}
    \end{subfigure}
    ~
    \begin{subfigure}
        \centering
        \includegraphics[width=0.45\textwidth,keepaspectratio]{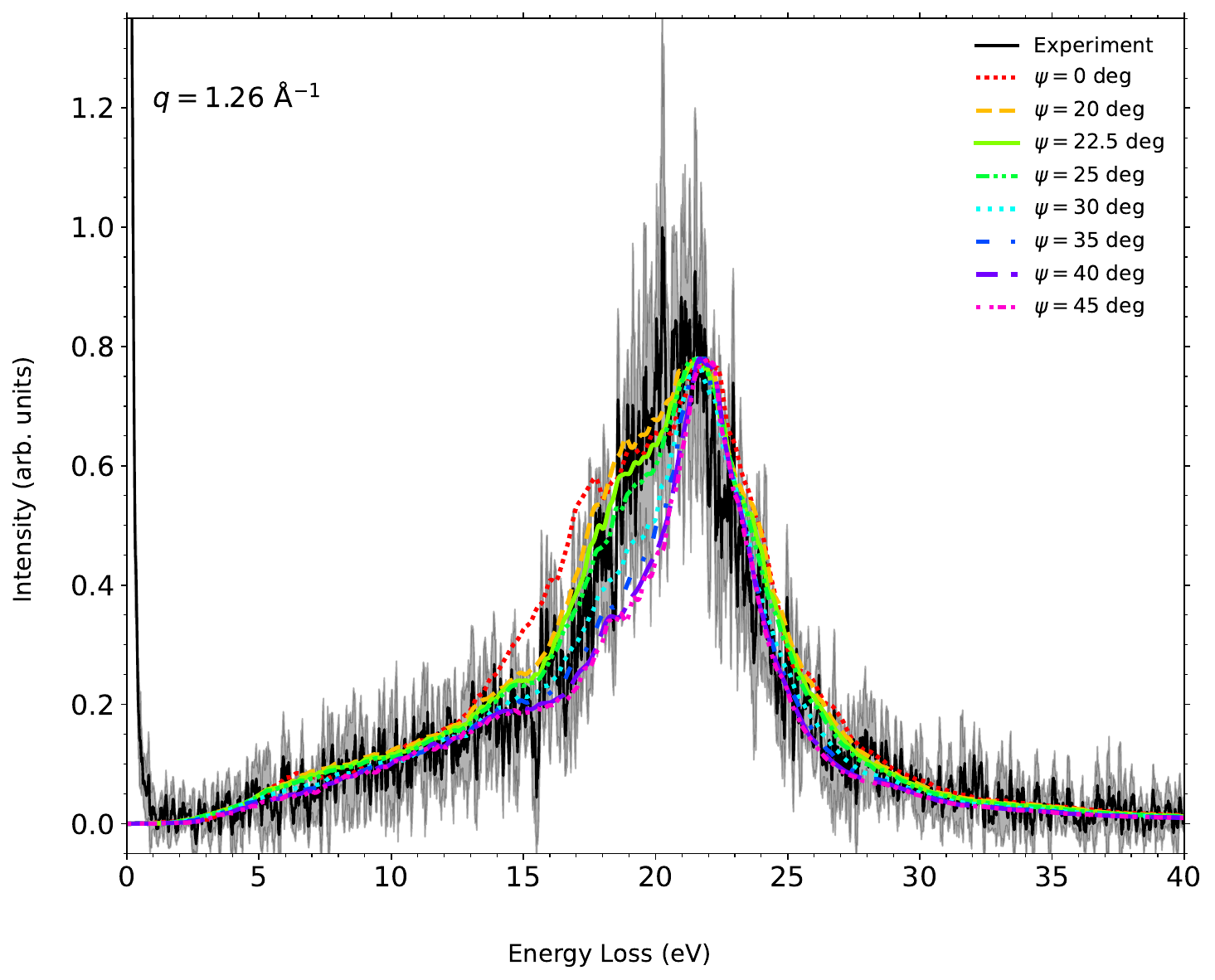}
    \end{subfigure}
    \caption{Comparison of the experimental data (black) to the TDDFT calculations of the DSF of Si at two central central scattering vectors of the DCA, but with outgoing wave vector rotated about the beam direction by an angle $\psi$ in Eq.~(\ref{eq:scatter_vector}).
    }
    \label{fig:TDDFT_Rotating}
\end{figure} 

First, we consider the predictions of TDDFT of the DSF along the principal [100], [110], and [111] directions.
The TDDFT-predicted DSFs along these directions are plotted with the experimental data at central scattering vectors $q=0.55 \, \textup{\AA}^{-1}$ and $1.26 \, \textup{\AA}^{-1}$ in Fig.~\ref{fig:TDDFT_Principal}. In these simulations, $q$-vector blurring has been accounted for by averaging over five simulations for different $q$ values that lie in the DCA range, but aligned along each of these three principal directions.
For $q=0.55 \, \textup{\AA}^{-1}$, the DSFs along the three directions look quite similar, but there are noticeable differences such as the higher intensity between 10--15~eV along the [100] direction compared to the other two. Slight differences are to be expected at this scattering vector as it lies just beyond the boundary of the first BZ (see Fig.~\ref{fig:Si_geom}~(a)) and so the structural details of the Si lattice are starting to be probed.
For $q=1.26\textup{\AA}^{-1}$, there are clearly substantial differences in the shapes of the spectra for the three different directions. This scattering vector now extends far out into reciprocal space, and so the DSF contains more information on the lattice structure along the particular scattering vector.
Notably, when the scattering vector is aligned along [100] -- corresponding to the normal of the Si sample -- there is a large bump centered around 15~eV which does not appear in the experimental data, indicating more detailed accounted of the experimental geometry is indeed warranted. 
Also note the superficially good agreement between the [111] direction and the experimental data: as seen in Fig.~\ref{fig:Si_geom}~(a), this $q$ value ($\theta=18.6^\circ$) does not lie near any [111]-like planes, so this apparent agreement is coincidental since the [111] direction was experimentally inaccessible.

Next, we consider the scattering vector as it would be aligned through the Si lattice. As described previously, for the experimental geometry shown in Fig.~\ref{fig:Setup}, the scattering vector through the lattice can be written as:
\begin{equation}
    \bm{q} = Q(\cos\Theta - 1, \, \sin\psi \sin\Theta , \, \cos\psi \sin\Theta ) \, .
    \label{eq:scatter_vector}
\end{equation}
Changing the angle $\psi$ is the equivalent of rotating around the circles in reciprocal space in Fig.~\ref{fig:Si_geom}~(a).
In Fig.~\ref{fig:TDDFT_Rotating}, we plot a comparison of the DSFs at $q=0.92 \, \textup{\AA}^{-1}$ (left) and $q=1.26 \, \textup{\AA}^{-1}$ (right) for different rotations $\psi$. Note that $q$-vector blurring has not been accounted for here. For $q=0.92 \, \textup{\AA}^{-1}$, increasing the angle $\psi$ from 0$^\circ$ to 20$^\circ$ results in the shoulder between 12--15~eV vanishing, a feature which is not observed in the experimental spectrum. At higher values of $\psi$, the peak becomes narrower.
For $q=1.26 \, \textup{\AA}^{-1}$, changing the value of $\psi$ has a substantial impact on the shape of the DSF as the scattering vector points along different directions through the crystal.

Both Fig.~\ref{fig:TDDFT_Principal} and Fig.~\ref{fig:TDDFT_Rotating} show that the predicted DSF from TDDFT is sensitive to the alignment of $\bm{q}$ through the crystal lattice, which is the expected behaviour. This also means that its predictions of the DSF are of S($\bm{q}$, $\omega$), not just the isotropic $S(q, \omega)$ which is typically used in approximating the form of the DSF.


\subsection{Comparison of theory to experiment}
\begin{figure}
    \centering
    \begin{subfigure}
        \centering
        \includegraphics[width=0.475\textwidth]{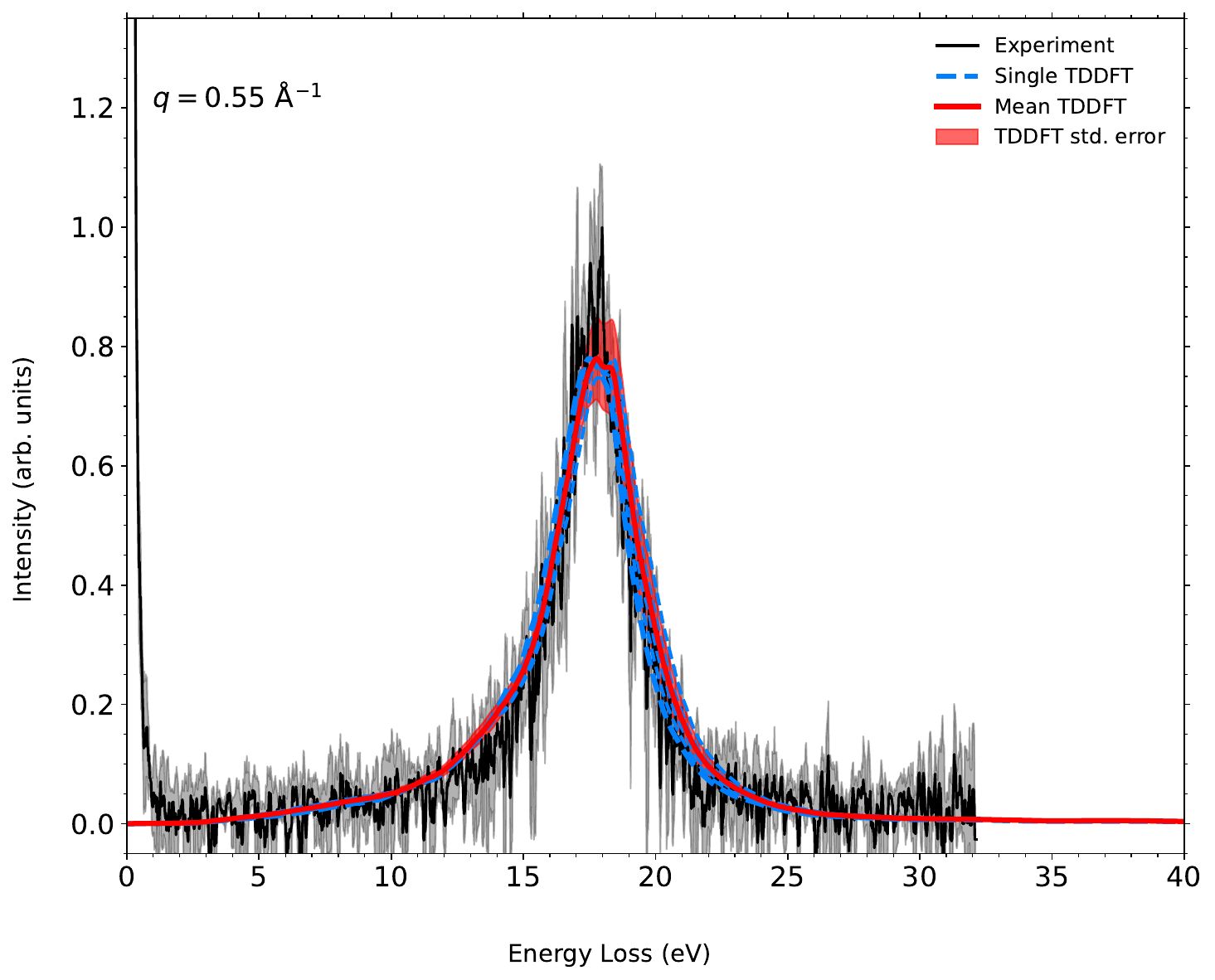}
    \end{subfigure}
    ~
    \begin{subfigure}
        \centering
        \includegraphics[width=0.475\textwidth]{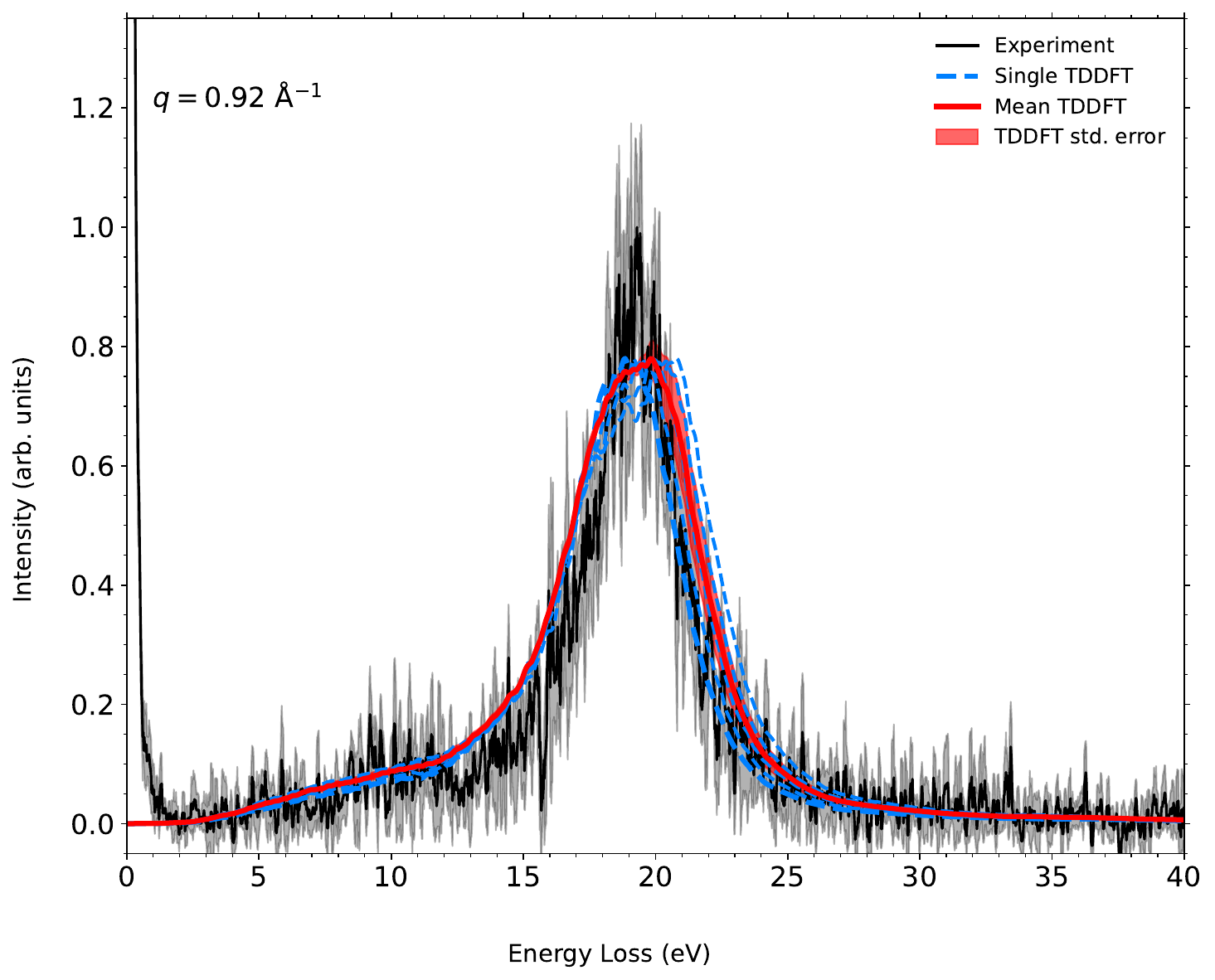}
    \end{subfigure}
    \vskip\baselineskip
    \begin{subfigure}
        \centering
        \includegraphics[width=0.475\textwidth]{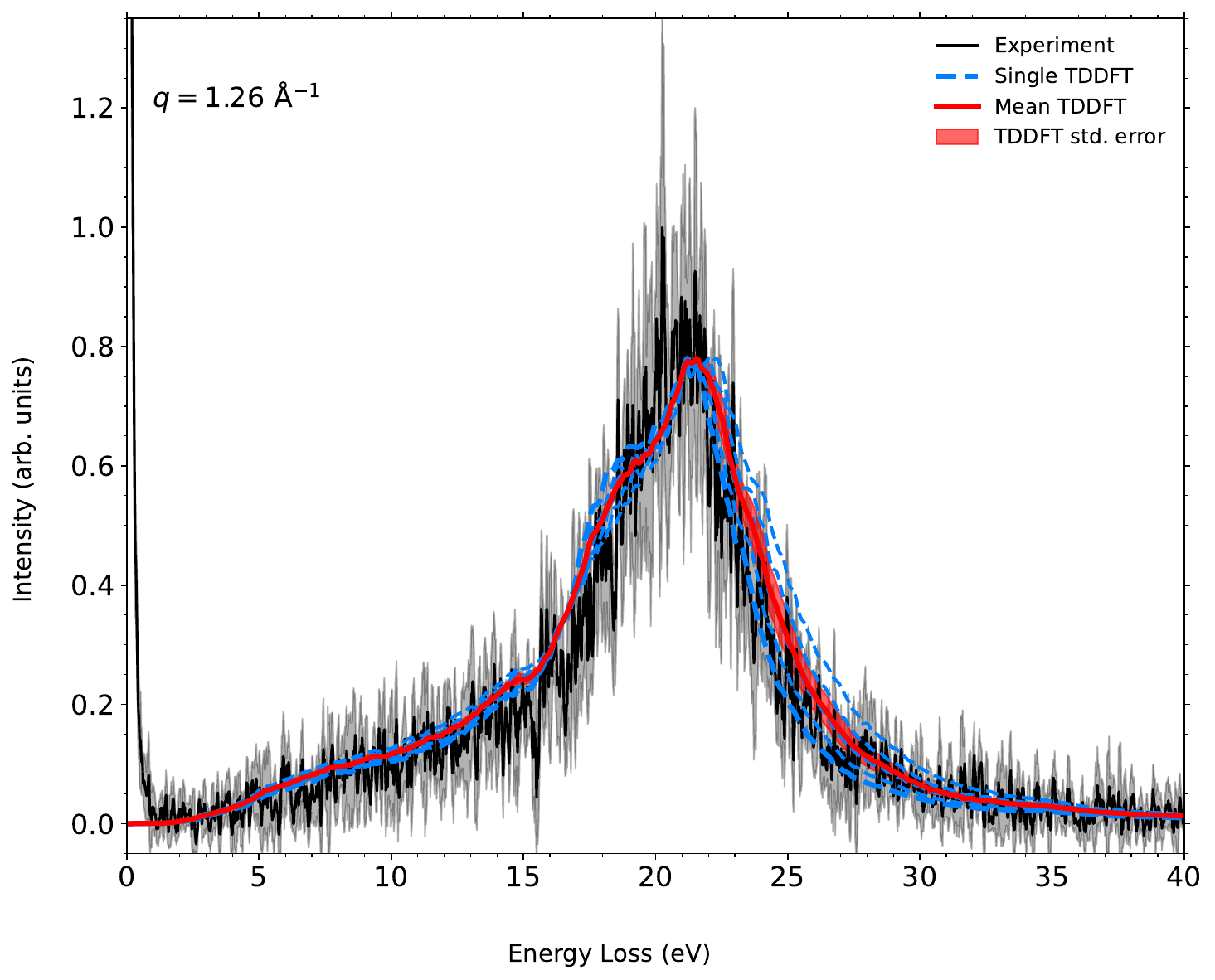}
    \end{subfigure}
    ~
    \begin{subfigure}
        \centering
        \includegraphics[width=0.475\textwidth]{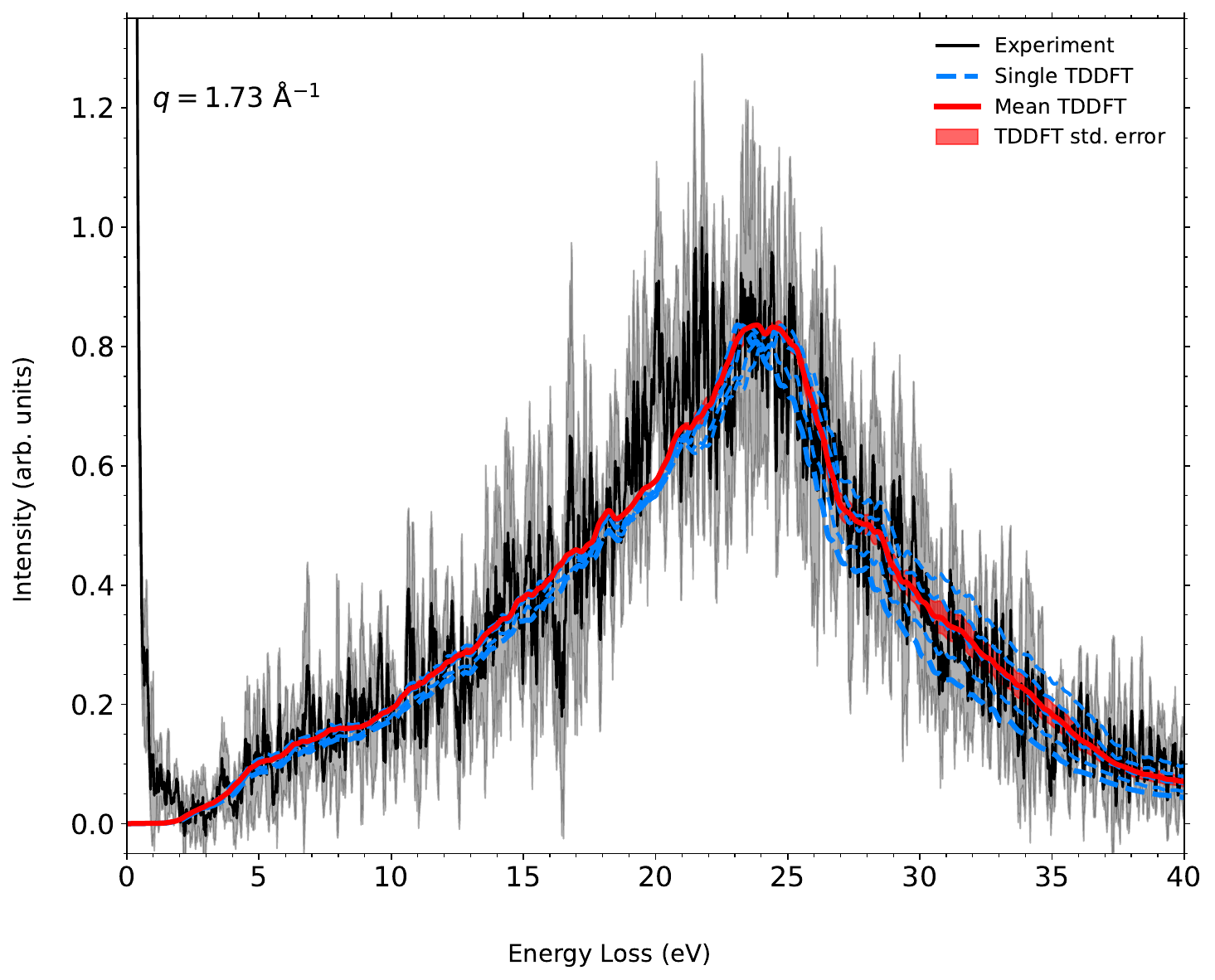}
    \end{subfigure}
    \caption{Comparison of TDDFT simulations of the DSF of Si, with $\psi=22.5^\circ$ to the experimental signal (black). For each scattering vector, five TDDFT simulations at different $q$ are shown as the blue-dashed lines in each plot. The average of these five simulations is shown as the red line, and is the final result to compare to the experimental spectra. The red bands indicate the standard error of the mean DSF. Each of the TDDFT lines are normalised to their maxima, and scaled to match the intensity of the experimental data.
    }
    \label{fig:TDDFT_final}
\end{figure}

As mentioned previously, in the present experiment the angle $\psi$ was not determined, but we do know it is the same for all measured scattering angles $\Theta$ as the Si wafer was not changed or reoriented. Instead, we compared the shape of the DSFs predicted by TDDFT for different values of $\psi$ to the experimental data at $q=1.26 \, \textup{\AA}^{-1}$ (since the theoretical DSF is most sensitive to the specific orientation scattering vector here), and concluded the best visual agreement came from using $\psi=22.5^\circ$. Here, this value of $\psi$ is now used to compare all TDDFT-predicted DSF to all the experimental spectra.
This of course caveats the quality of the benchmarking done here as TDDFT is being used to determine an unknown parameter; however, given the clear sensitivity of TDDFT to the scattering geometry, that the geometry is otherwise well-defined, and the subsequent agreement we now observe between TDDFT and experiment, we remain confident in the quality of the following benchmarking.

It is worth discussing now in detail how to interpret the finite the size of the spectrometer for the simulations. As mentioned already, the (masked) DCA covers a range of outgoing wave vectors $\bm{q_2} \pm \Delta \bm{q_2}$ around the central wave vector $\bm{q_2}$. This gives a range of scattering vectors $\bm{q} \pm \Delta \bm{q}$ that are measured on the spectrometer, here this is around $\Delta q \sim \pm 0.1 \textup{\AA}^{-1}$. But, as already mentioned in the discussion of the dispersion, this cannot be treated as an uncertainty on $\bm{q}$ because it is not. Instead, it means the measured spectra contain scattering from multiple scattering vectors. The way in which these different scattering spectra contribute to the overall spectrum  will depend on the particular spectrometer geometry. Here, the horizontal slit over the DCA means the distribution of die is uniform, so that it reflects the scattering spectra from the different $\bm{q}$ vectors it covers uniformly to each energy bin. Therefore, the measured spectrum will be the uniform average of all scattering vectors covered by the DCA.

To simulate the $q$-vector blurring, five TDDFT simulations are performed for a uniform set of $\Theta$ in Eq.~(\ref{eq:scatter_vector}) within the range of the DCA at the central scattering angle.
These results are plotted in Fig.~\ref{fig:TDDFT_final}, along with the five individual calculations. For the plotting, each of the TDDFT lines are normalised to their maxima, and scaled to match as closely as possible the shape of experimental curves. To be clear, for the averaged DSF the contributing DSFs were not normalised in any way before averaging.

In general, we find very good agreement between the average DSFs predicted by TDDFT and the experimental data. As with the Al data reported in Ref.~\cite{Gawne_2024_Ultrahigh}, we observe that accounting for $q$-vector blurring is still important in Si.
As the scattering angle increases, we observe the effect of the $q$-vector blurring becomes more substantial, and the differences between the individual curves grow, particularly in the high energy loss wings of the spectra. It also means none of these single spectra can be used to model the experimental signal.
This demonstrates that treating the finite coverage of the spectrometer as an uncertainty bar on the central $\bm{q}$ is actually detrimental to interpreting the spectrum, and that only by accounting for the geometry of the spectrometer in the simulated spectrum can one reproduce the scattered spectrum.
 
We note that our simulated spectra use a fixed Lorentzian broadening, and still give good agreement. This in contrast to other reported simulations of the Si DSF which instead claim energy-dependent broadening is required. For example in Ref.~\cite{Weissker_2010_Dynamic}, the energy-dependent broadening is introduced either in the evaluation of the density response function $\chi_0$ as an energy-dependent lifetime (EDLT) broadening; and alternatively as a post-processing step by applying a Lorentzian with an energy-dependent width as a kernel to the TDDFT results, although the authors state the latter fails to adequately reproduce their experimental data.
The authors claim an energy-dependent broadening is required to reduce the sharpness of features from the unmodified TDDFT results, which were not observed in their experiment~\cite{Weissker_2010_Dynamic}. However, we note that the use of EDLT still produces TDDFT data that has notable differences compared to their measured spectra, such as the peak intensity and the width of their calculated DSFs. To our understanding, the authors did not account for $q$-vector blurring from the finite size of their spectrometer, which we expect would help to smooth out the sharp features from their TDDFT calculations, as well as helping to broaden the high energy loss wings of their DSFs, which seemed to be underestimated compared to their experimental spectra even with EDLT.
We note that the good agreement we achieve here is also only at the level of ALDA. We therefore question whether energy-dependent broadening is actually needed to explain the broadening of the DSF, and suggest that $q$-vector blurring probably explains the bulk of the differences between TDDFT results and experiment.

That being said, we still observe some differences between the TDDFT spectra presented here and the experimental data. In particular, for $q=0.92 \, \textup{\AA}^{-1}$ the TDDFT spectra seems to overestimate the width of the spectrum.
We note that while Fig.~\ref{fig:TDDFT_Rotating} indicates a narrower peak would come from using $\psi \ge 30^{\circ}$, this would result in even more pronounced discrepancies for $q=1.26 \, \textup{\AA}^{-1}$, so there is a constraint how large the angle $\psi$ could be while maintaining consistent results between the different scattering angles.
One alternative explanation is that we are only averaging five DSFs, which show quite substantial changes, and increasing the number of single TDDFT calculations may improve the averaging. Still, the standard error of these average DSFs suggest the averaging is already relatively near the actual value. A related possibility comes from the fact that we only consider $q$-vectors by varying $\Theta$ in the polar direction, but the DCA covers a finite range of angles in the azimuthal direction too. These were neglected as adequate sampling would require a very large number of simulations even for a single $q$.
However, we estimate that the effect of including these additional scattering vectors would be small: since the set of scattering vectors with the same scattering angle $\Theta$ and magnitude form a cone around $\bm{q_1}$, the effect of the $q$-blurring in the azimuthal direction is to add an additional angle $\phi$ to the angle $\psi$, so the scattering vectors are therefore:
\begin{equation}
    \bm{q} = Q(\cos\Theta - 1, \,  \sin(\psi+\phi)\sin\Theta, \, \cos(\psi+\phi) \sin\Theta) \, .
    \label{eq:full_scatter_vector}
\end{equation}
In the azimuthal direction, the slit limits the angular coverage $\phi\sim\pm2.5^\circ$. As seen in Fig.~\ref{fig:TDDFT_Rotating}, the shape of the DSFs for $\psi=20^\circ$, $\psi=22.5^\circ$ (the estimated nominal), and $\psi=25^\circ$ look quite similar.
Therefore, while contributions from these additional directions may affect the overall shape of the averaged DSF, we estimate that the average spectrum presented here would probably be quite similar even if these contributions were included.
An additional possibility comes from the fact that we are only treating Si on the level of ALDA. This approximation explicitly treats exchange and correlation on the level of a uniform electron gas, which semi-conducting Si with its covalent bonds is certainly not. Climbing further up the ``Jacob's ladder'' of exchange-correlation approximations~\cite{martin_reining_ceperley_2016, Perdew_Jacobs_Ladder, Tao_Jacobs_Ladder}, for example to account for density gradients in the exchange-correlation, may improve the quality of the simulations further. Nevertheless, ALDA manages to estimate the shape of the DSF satisfactorily.


\section{Summary and Discussion}\label{s:summary}

We have presented ultrahigh resolution data for the inelastic scattering of single crystal silicon at a number of scattering angles. Unlike previous experiments, where the scattering vector is kept aligned to a particular orientation of the crystal~\cite{Stiebling_1978_Dispersion,Chen_1980_Bulk,Schuelke_1995_Dynamic,Weissker_2010_Dynamic}, here we fix the normal direction of the crystal along the incident beam axis and allow the orientation of the scattering vector to change through the crystal.
While doing this makes it difficult to infer the quadratic dispersion of the Si plasmon, it on the other hand draws out the complex strong-geometry dependent behaviour of the inelastic scattering signal, and presents a challenging benchmark to ascertain if time-dependent density functional theory can accurately model this complex geometry.

In TDDFT simulations, we observe strong dependencies of the DSF on the particular orientation of the scattering vector through the lattice. This is interpreted from the perspective of reciprocal lattice space, with the different directions and length of the scattering vectors probing nearby lattice points -- and therefore structures -- of the crystal lattice. Even rotating the lattice about its normal is shown to result in substantial changes in the shape of the DSF due to the distribution of neighbouring reciprocal lattice vectors.

In comparison to experiment, we find TDDFT is able to accurately model the shape of the DSF if $q$-vector blurring is accounted for. This also comes with the caveat that the specific orientation of the Si sample needed to be inferred from TDDFT as it could not be measured in experiment. However, given the sensitivity of TDDFT to the scattering geometry, we remain confident in the quality of the benchmarking.
While there are some minor discrepancies between TDDFT and experiment, we identify a number of potential improvements to the simulation scheme that might reduce them. Regardless, our results suggest that modified TDDFT schemes involving energy-dependent broadening~\cite{Weissker_2010_Dynamic} may be unnecessary, as the smoothing out and broadening of features seems to be adequately explained by $q$-vector blurring.

Additionally, we note that the experimental data presented here was collected at a much faster rate than in Ref.~\cite{Gawne_2024_Ultrahigh} in a material that scatters more weakly than the other dataset. In both cases the beam performance was unsatisfactory, and in normal conditions substantially more photons would be incident on the target.
We have therefore demonstrated that, even with less than ideal conditions, this experimental setup~\cite{Gawne_2024_Ultrahigh} can be used to efficiently collect high quality spectra. We suggest here that an initial fast collection rate run could be used to identify the location of important features, and then longer more focused scans over the identified features to acquire the desired level of signal-to-noise ratio, would together improve the overall collection rate in an experiment. We therefore have an optimistic outlook on the utility of this ultrahigh resolution setup in a broader range of experimental scenarios, such as the application of the recent model-free thermometry approach~\cite{Dornheim_T_2022,Dornheim_T2_2022} at comparably low temperatures of $T\sim1\,$eV.

Finally, we conclude that the results presented here give confidence in the predictive capability of TDDFT in a single crystal semi-conductor, while the results reported in Ref.~\cite{Gawne_2024_Ultrahigh} demonstrated its predictive capability for a simple metal. This is particularly exciting as recently reported TDDFT simulations of Si and Al undergoing isochoric heating showed a number of unexpected behaviours~\cite{Moldabekov_2024_Excitation}. Notably, for a small amount of heating, the plasma frequency in Al and in Si along the [100] direction are predicted to lower rather than increase. On the other hand, along the [111] direction in Si, plasmon lowering is not predicted, but thermally-induced excitations do appear in the DSF near zero energy loss. The fact that TDDFT can accurately simulate geometric effects in the DSF is crucial to having confidence in its predictions of different DSF behaviours along the different axes.
Additionally, it has also been recently reported that TDDFT predicts complex geometric- and thermal-dependencies on the DSF of fcc copper (Cu)~\cite{moldabekov2024ultrafast}. However, as a $d$-band metal, Cu presents new challenges to TDDFT that require further benchmarking from what has so far been achieved.
In all cases, these changes to the DSF typically occur over very small energy scales, and necessitate the need for an ultrahigh resolution setup, for example to accurately infer plasmon lowering from the plasmon dispersion, and to distinguish the thermally-induced excitations from the quasi-elastic peak~\cite{Moldabekov_2024_Excitation}.
In conclusion, the successful application of this setup both to make ultrahigh resolution measurements and to confirm the predictive power of TDDFT in multiple systems opens the door to exciting investigations of systems in more exotic conditions.

\section*{Data Availability Statement}
The original datasets can be found here and are available upon reasonable request: doi:10.22003/XFEL.EU-DATA-003777-00.

\section*{Conflict of Interest Statement}
The authors have no conflicts of interest to disclose.

\section*{Acknowledgements}
We acknowledge the European XFEL in Schenefeld, Germany, for provision of X-ray free-electron laser beamtime at the Scientific Instrument HED (High Energy Density Science) under proposal number 3777 and would like to thank the staff for their assistance. The authors are grateful to the HIBEF user consortium for the provision of instrumentation and staff that enabled this experiment. 

This work was partially supported by the Center for Advanced Systems Understanding (CASUS), financed by Germany’s Federal Ministry of Education and Research (BMBF) and the Saxon state government out of the State budget approved by the Saxon State Parliament. 
This work has received funding from the European Union's Just Transition Fund (JTF) within the project \emph{R\"ontgenlaser-Optimierung der Laserfusion} (ROLF), contract number 5086999001, co-financed by the Saxon state government out of the State budget approved by the Saxon State Parliament.
This work has received funding from the European Research Council (ERC) under the European Union’s Horizon 2022 research and innovation programme
(Grant agreement No. 101076233, "PREXTREME"). 
Views and opinions expressed are however those of the authors only and do not necessarily reflect those of the European Union or the European Research Council Executive Agency. Neither the European Union nor the granting authority can be held responsible for them. Computations were performed on a Bull Cluster at the Center for Information Services and High-Performance Computing (ZIH) at Technische Universit\"at Dresden and at the Norddeutscher Verbund f\"ur Hoch- und H\"ochstleistungsrechnen (HLRN) under grant mvp00024. 
%

\section*{References}

\bibliographystyle{unsrt}
\bibliography{ref}

\end{document}